# Subtle Contact Nuances in the Delivery of Human-to-Human Touch Distinguish Emotional Sentiment

Shan Xu, Chang Xu, Sarah McIntyre, Håkan Olausson, and Gregory J. Gerling

*Abstract*—We routinely communicate distinct social and emotional sentiments through nuanced touch. For example, we might gently hold another's arm to offer a sense of calm, yet intensively hold another's arm to express excitement or anxiety. As this example indicates, distinct sentiments may be shaped by the subtlety in one's touch delivery. This work investigates how slight distinctions in skin-to-skin contact influence both the recognition of cued emotional messages (e.g., anger, sympathy) and the rating of emotional content (i.e., arousal, valence). By self-selecting preferred gestures (e.g., holding, stroking), touchers convey distinct messages by touching the receiver's forearm. Skin-to-skin contact attributes (e.g., velocity, depth, area) are optically tracked in high resolution. Contact is then examined within gesture, between messages. The results indicate touchers subtly, but significantly, vary contact attributes of a gesture to communicate distinct messages, which are recognizable by receivers. This tuning also correlates with receivers' arousal and valence. For instance, arousal increases with velocity for stroking, and depth for holding. Moreover, as shown here with human-to-human touch, valence is tied with velocity, which is the same trend as reported with brushes. The findings indicate that subtle nuance in skin-to-skin contact is important in conveying social messages and inducing emotions.

*Index Terms*—Social touch, haptics, perception, emotion communication.

## I. Introduction

Human-to-human touch is essential to social communication, particularly in expressing emotion. For example, those in intimate relationships convey love and sympathy, often preferring touch over facial expressions, body postures, or movements [1]. Social and affective touch is also critical in cognitive development throughout infancy and childhood, providing emotional support and forming social bonds [2]. Moreover, works are now indicating that social meaning is readily identified from touch alone [3]–[6].

Certain touch interactions may underlie how we communicate social and emotional sentiment. To understand contact deployed in human-to-human touch, prior efforts have used human observers to annotate a toucher's gestures, contact duration, and contact intensity [3], [4], [7]. While touchers regularly vary their gestures to convey different intentions, a more interesting observation is that they often reuse the same gesture to communicate multiple, distinct messages. For example, stroking might convey both love and sadness, while shaking might convey both happiness and anger [3], [4]. To

This work was supported in part by the National Science Foundation (IIS-1908115) and the National Institutes of Health (R01NS105241).

S. Xu, C. Xu, and G. J. Gerling are with the School of Engineering and Applied Science, University of Virginia, Charlottesville, VA, USA. S. McIntyre and H. Olausson are with the Center for Social and Affective Neuroscience (CSAN), Linköping University, Sweden. The corresponding author is G. J. Gerling (phone: 434-924-0533; e-mail: gg7h@virginia.edu).

capture detail finer than possible with a proctor, computerized tracking systems have been introduced. Such systems utilize sensors, cameras, and electromagnetic trackers [6], [8]–[12] to quantitatively capture pressure and positional attributes that underlie human-to-human contact. Certain attributes (e.g., tangential velocity, indentation depth, contact area) seem to be important in distinguishing social messages (e.g., love, calm, happiness) and may be intuitively understood between participants without training [11].

However, we still do not understand whether receivers actually perceive emotions from these touch expressions, as opposed to intuitively discriminating and associating touch expressions as 'codes' with the messages. Classic theory on emotion dictates two dimensions are at play, i.e., valence and arousal, which refer to the pleasantness and the intensity of emotion, respectively [13]. As a point of comparison, under the brush stroke to the forearm and hand, changes in velocity govern pleasantness, with an optimal speed range around 1-10 cm/s [14]. In other experimental paradigms, changes in valence and arousal have been examined when participants observe images of social and non-social touch [15] and in physical interactions with robot hands [16]. Here, in human-to-human interaction, we consider whether slight changes in someone's touch delivery alter the valence and arousal perceived by receivers. Different from well-controlled machine-delivered touch, in natural human touch, multiple contact attributes change in unique combinations, and frequently, over the course of a message's delivery. In this context, quantifying the impact of subtle contact changes is difficult, due to the complexity inherent in measuring skin-to-skin contact, as even thin barriers alter perception [17].

Herein, by developing an interference-free visual tracking system to quantify contact attributes, we investigate if slight distinctions in skin-to-skin contact tune both receivers' recognition of a cued message (e.g., anger, sympathy) and their ratings of its emotional content (i.e., arousal, valence).

## II. Methods

A high-resolution 3D visual tracking system is developed to quantify physical interactions with a receiver's forearm for seven contact attributes: contact area, indentation depth, contact duration, absolute velocity, and three orthogonal velocity components. A unique experimental paradigm is used wherein touchers convey distinct cued emotional messages to the receiver's forearm by use of self-selected gestures. In particular, touchers deliver one of seven messages [3]–[6], using one of three commonly employed gestures for that message [3]–[6]. Upon a message's delivery, receivers report the message they recognize and rate its valence and arousal.

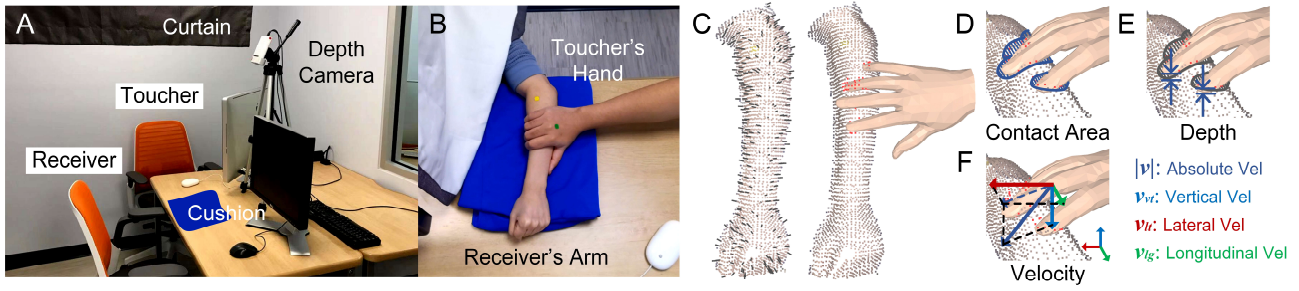

*Figure 1.* Experimental setup and hand-to-forearm contact measurement. (A) Touchers and receivers are separated by an opaque curtain and without verbal communication. Instructions and questions are displayed and recorded by computers in front of participants. The depth camera (Kinect) was set in between to capture hand-arm interactions. (B) A snapshot taken by the depth camera during the experiments. The toucher was delivering a cued emotional message to the receiver's forearm. (C) An example 3D forearm point cloud and hand mesh. On the forearm, black line segments denote a normal vector per point, and red points illustrate the contact region. Six time-series contact attributes include: (D) contact area as the overall area of the contact region on forearm, (E) indentation depth as the average indentation of all contacted hand points relative to the forearm surface, (F) absolute value of hand velocity and its three orthogonal components. The vector of spatial hand velocity is derived from the position of the middle metacarpophalangeal joint. Longitudinal velocity is along the direction from receiver's elbow to wrist. Vertical velocity follows the vertical direction pointing up. Lateral velocity is perpendicular to the other two velocities pointing left.

## A. Experimental Procedures

*Setup*: As illustrated in Fig. 1A, an opaque curtain was set between the toucher and receiver to inhibit visual cues and they were instructed not to speak to one another. A cushion was set on the toucher's side of the table upon which the receiver's forearm could rest. All experimental instructions and participant responses were delivered and recorded by a graphical user interface on either participant's computer. To measure physical contact interactions, a time-of-flight depth camera (Azure Kinect, Microsoft, USA) was mounted on a tripod in front of the cushion and oriented towards the cushion.

*Cued message and gesture stimuli*: Adopted from prior works [3]–[6], seven cued emotional messages were provided with each message associated with three commonly used gestures (Table I). Among those cued messages, anger, happiness, and fear are universally recognizable through facial, vocal, and touch expressions [5]. Gratitude and sympathy are prosocial and easily communicated by touch [3]–[5]. Attention and calm are interpreted by touch significantly better than chance [6], [11]. Among the gestures, holding and squeezing were combined as one choice given their comparable poses and movements. Similarly, hitting and tapping were combined, but only for the expression of anger.

*Participants*: The study was approved by the Institutional Review Board at the University of Virginia. For each experimental group consisting of three participants, at least one male and one female participant were recruited. Five groups were assembled, resulting in 15 total participants (8 male and 7 female, mean age = 23.9, SD = 4.6), and all provided written informed consent.

*Procedures*: Each group completed four experimental sessions. For two of the sessions, one male participant from that group was assigned as the toucher, while the other two participants were assigned as the receivers in each session, respectively. For the other two sessions, one female participant was assigned as the toucher and the other two participants were assigned as receivers. The four sessions were randomized per group, with a 20-minute break between sessions. In each session, seven messages were communicated with each message repeated six times. The 42 touch communication trials were performed randomly. Therefore, four sessions of 42 trials each were completed by each of the five experimental groups, for a total of 840 trials. Neither cued messages nor gesture stimuli were revealed to the participants before experiments.

In each trial, one message word was displayed to the toucher on the screen along with the three gestures. The order of displayed gestures was randomized per trial. The toucher was allotted 5 seconds to choose one gesture for that message and report it on the user interface. Then, the toucher delivered the message, by touching the receiver's left forearm between the elbow and wrist, using the right hand. Only the chosen gesture could be used in that trial, a combination of multiple gestures was not allowed. For the same message in different trials, producers were free to reuse the same gesture or change to another. No constraints or instructions were given for delivering a gesture. Touchers could use whichever patterns of contact they deemed appropriate for that gesture, with any duration or repetition within a trial for more natural contact interactions. After the conclusion of contact, the toucher clicked a button to inform the receiver. The receiver was then asked to select the message recognized from contact, among a list of the same seven words in a random order. The valence and arousal states perceived from contact were also rated by receivers, using the Self-Assessment Manikin (SAM) method with 9 levels from unpleasant to pleasant and from calm to excited, respectively [18].

## B. Physical Contact Attributes

An interference-free 3D motion-tracking pipeline was developed based on a high-resolution Azure Kinect camera. Toucher's hand and receiver's forearm were tracked separately but simultaneously within the same camera coordinate (Fig. 1B). The shape of the forearm was extracted using a region growing segmentation algorithm to cluster all arm points into one group. Considering that the forearm is often occluded by the toucher's hand, its shape was captured before each trial. During the trial, the position of arm shape was updated continuously via a color marker on the arm (Fig. 1B). The normal vector of each arm point was derived for contact detection and quantification (Fig. 1C).

TABLE I.     CUED MESSAGES AND ASSOCIATED GESTURES

| Message | Gestures | | |
|---|---|---|---|
| Anger (Ag) | Hit/Tap | Hold/Squeeze | Shake |
| Attention (At) | Tap | Shake | Hold/Squeeze |
| Calm (C) | Hold/Squeeze | Stroke | Tap |
| Fear (F) | Hold/Squeeze | Shake | Tap |
| Gratitude (G) | Hold/Squeeze | Shake | Tap |
| Happiness (H) | Shake | Tap | Stroke |
| Sympathy (S) | Stroke | Tap | Hold/Squeeze |

The hand tracking procedure was developed based on a monocular hand tracking algorithm, which is robust to occlusions and object interactions [19]. Twenty-one hand joints and one solid-color marker on the back of the hand (Fig. 1B) were first detected from color images. Then, 3D positions and rotations of hand joints were predicted relative to the hand coordinate using a convolutional neural network and inverse kinematic network. By merging depth information, the detected 2D marker was transformed into 3D following the camera projection model. The spatial hand position expressed in the camera coordinate was next derived according to the 3D marker. Finally, by fitting the open-source MANO hand model [19] to the spatial pose and position of 21 joints, a 3D mesh of the hand was animated in real-time following the movement and gesture of the toucher's hand.

Contact interactions between the toucher's hand and the receiver's forearm were detected and measured in a point-based manner. As shown in Fig. 1C-D, skin contact was detected if any vertex point of the hand mesh was captured beneath the forearm surface. Upon contact, six time series contact attributes, i.e., contact area, absolute contact velocity, decomposed longitudinal, lateral, and vertical velocities, and indentation depth, were derived per trial from the initiation of contact to its conclusion. To obtain the scalar value of those time series attributes, the mean value was extracted as representative. The seventh contact attribute, i.e., contact duration, was scalar and defined as the overall duration for which contact was detected per trial. Contact area was calculated as the sum of unit area of all contacted arm points (Fig. 1D). Indentation depth was computed as the average distance of all contacted hand points relative to the arm surface (Fig. 1E). Absolute contact velocity was the norm of hand velocity vector derived from spatial hand position. The middle metacarpophalangeal joint was chosen to represent hand position, given its robustness over diverse set of gestures and interactions. Inspired by prior measurements of normal and tangential velocities with direction information [6], [10], [11], we further decomposed the spatial velocity vector into three orthogonal components. Longitudinal velocity was aligned with the arm direction pointing from elbow to wrist. Vertical velocity was in the vertical direction pointing upward. Lateral velocity was perpendicular to the other two directions pointing to the internal side of the forearm (Fig. 1F).

### C. Data Analysis

Out of the 840 trials, 823 were analyzed. A few trials were either skipped by participants or not properly tracked. Statistical and machine learning analyses were used to examine the functional roles of contact attributes, regarding the recognition of delivered messages and the ratings of valence and arousal states. Three main analyses were performed, as ultimately associated with Figs. 2, 3, and 5. First, the proportions of messages delivered by touchers, and recognized by receivers, using one touch gesture were analyzed, separately. The message recognition matrix was then calculated per touch gesture.

Second, Mann–Whitney U tests were performed to evaluate differences in contact attributes across messages given the same gesture. Since receivers might interpret messages different from those intended, this test was conducted per gesture for both delivered and recognized messages. Benjamini-Hochberg correction was applied to the multiple testing for each attribute within a gesture by controlling the false discovery rate. In cases where a message was recorded for less than 15 trials with a gesture, it was excluded from analysis due to the low statistical power. Next, in order to identify contact attributes that are salient in helping recognize messages, the importance of individual attributes was derived according to their contribution in predicting messages. To do so, among trials where messages were correctly recognized, a random forest algorithm was used to classify messages based on all attributes. The importance of attributes as predictors was derived based on Gini impurity. Due to the stochastic nature of the classifier, importance was calculated by averaging 100 repetitions per gesture.

Third, to evaluate valence and arousal, ratings per gesture of each message were grouped and averaged. Then, ratings for touch-delivered messages were compared with the ratings collected from the written word stimuli with the same words [20]. Next, to examine the relationship between contact attributes and valence and arousal ratings, linear and quadratic regressions were used. Spearman's rank correlation analysis and F test were applied to linear and quadratic relationships respectively and were corrected by Benjamini-Hochberg method per gesture.

### III. RESULTS

#### A. Same touch gesture communicates multiple messages

Touchers were able to use the same gesture to deliver multiple cued messages. Touch receivers were able to identify the similar sets of messages delivered by touchers, with discrepancies in only the relative proportions of delivered and recognized messages (Fig. 2A). In specific, touchers deployed stroking to deliver calm (32.9%), sympathy (41.1%), and happiness (24.7%). Touch receivers identified the same three messages, though calm (41.8%) was recognized in a relatively greater proportion than it was delivered. With the shaking gesture, touchers used it for fear (32%), happiness (29%), gratitude (21%), and attention (10%). Likewise, receivers perceived shaking as fear (39%) and attention (33.5%), with the other two messages recognized less often. With holding and tapping gestures, the proportions between touchers and receivers were nearly equivalent.

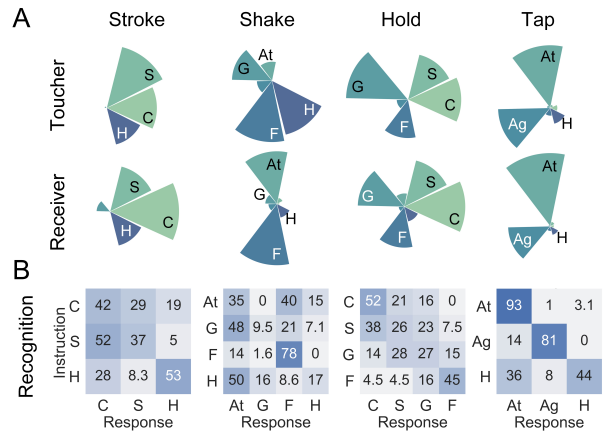

*Figure 2.* Messages communicated per gesture and recognition rates. (A) The proportion of messages delivered by touchers and recognized by receivers, respectively. Seven cued messages were listed in Table I. (B) The receiver's recognition rates of delivered messages per gesture. Within each cell, the value and color redundantly show the percentage of the recognized message. As noted, a few rarely conveyed messages were excluded, so the sum of each row may not total 100.

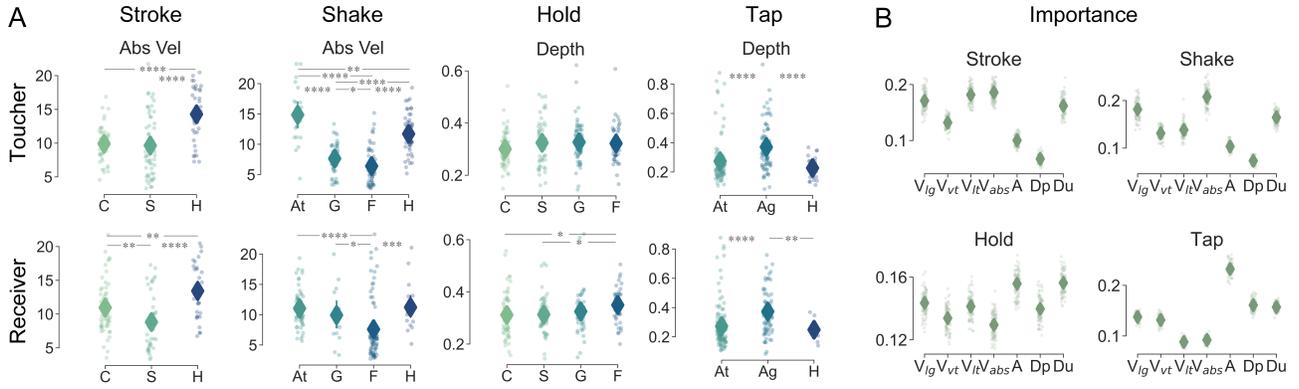

*Figure 3.* (A) Contact attributes of messages delivered by touchers (top row) and recognized by receivers (bottom row). One attribute that well distinguishes messages, i.e., absolute velocity (cm/s) or indentation depth (cm), is displayed per gesture. Diamonds denote means; points denote trial data. Error bars denote 95% confidence intervals and *$p < 0.05$, **$p < 0.01$, ***$p < 0.001$, ****$p < 0.0001$ are derived by paired-sample Mann–Whitney U tests after Benjamini-Hochberg correction. (B) Importance of contact attributes derived by their contribution in classifying successfully recognized messages. $V_{lg}$: longitudinal velocity (cm/s), $V_{vt}$: vertical velocity (cm/s), $V_{lt}$: lateral velocity (cm/s), $V_{abs}$: absolute velocity (cm/s), A: contact area (cm²), $Dp$: indentation depth (cm), $Du$: duration (s). Diamonds denote means; error bars denote 95% confidence intervals; points denote importance values by 100 repetitions of classification.

Recognition of messages is shown as a confusion matrix per gesture (Fig. 2B). When stroking was used, messages of calm and sympathy were apt to be confused, with happiness more easily recognized (53%). With the shaking gesture, fear was much more effectively communicated. With the holding gesture, calm (52%) and fear (45%) were recognized with higher accuracy relative to sympathy and gratitude. The tapping gesture was readily capable of getting one's attention (93%) or delivering anger (81%).

### B. Contact attributes change per message within gesture

To convey different messages, touchers slightly varied their contact attributes, which could be distinguished by receivers. In specific, Fig. 3A shows the distribution of contact attributes across messages per gesture, as conveyed by touchers and recognized by receivers, respectively. For example, with the stroking gesture, hand movements with significantly higher velocity were commonly associated with happiness. For the shaking gesture, fear was distinct from other messages with significantly lower velocity, resembling more closely a trembling motion than a vigorous shake. Given minimal hand motion with the holding gesture, only small changes in contact attributes were detected, yet fear could still be distinguished by its significantly greater indentation depth. With tapping, anger was also made distinct by utilizing significant greater indentation depth.

Practically speaking, the relative magnitudes of those contact attributes align with the expected practice of the cued messages, e.g., quicker stroking in happiness than sympathy and tighter holding in fear than calm. Indeed, a comparison of the magnitudes of the contact attributes between touchers and receivers shows similar trends, i.e., comparing down the column in Fig. 3A. That said, slight disparities still exist, such as attention expressed by shaking, with lower velocity by receivers. To provide the overall contact profile of cued messages recognized by receivers, the distribution of all attributes is shown (Fig. 4). Several contact attributes significantly distinguish recognized messages.

### C. Relative importance between contact attributes

Some contact attributes held more importance in distinguishing the recognized messages than others (Fig. 3B).

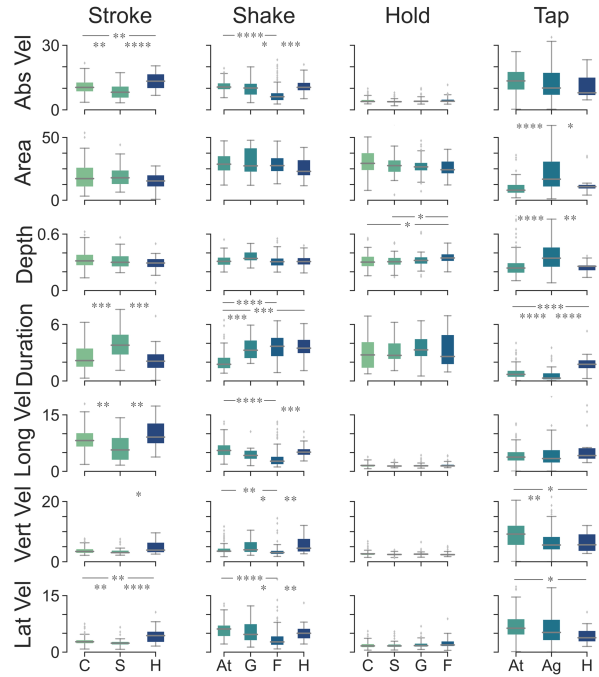

*Figure 4.* Absolute measurements per contact attribute, message, and gesture. Points denote trial data. *$p < 0.05$, **$p < 0.01$, ***$p < 0.001$, ****$p < 0.0001$ are derived by paired-sample Mann–Whitney U tests after Benjamini-Hochberg correction.

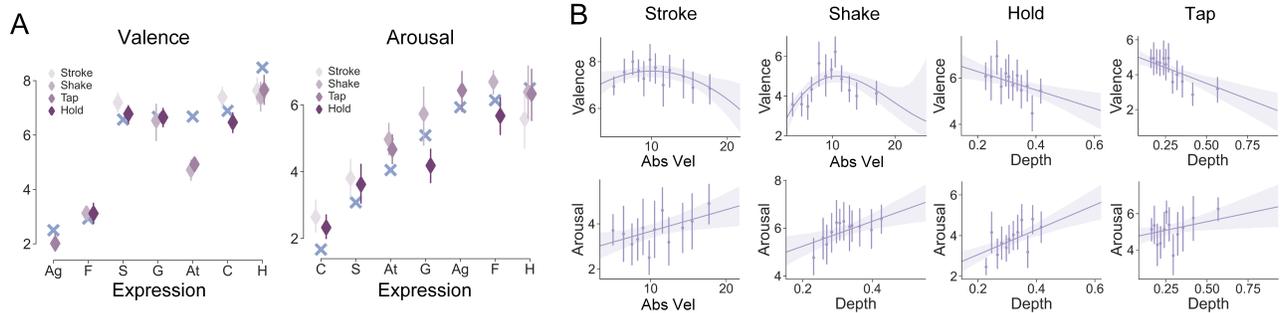

*Figure 5.* (A) Valence and arousal ratings are similar between gestures per message. Diamonds denote means and error bars denote 95% confidence intervals. Blue cross marks represent valence and arousal ratings retrieved from dataset using written word stimuli [15]. (B) Contact attributes distinctly tune arousal and valence ratings within a gesture. Absolute velocity (cm/s) and indentation depth (cm) exhibit clear trends. In each plot, trial data were binned into 12 evenly-sized groups. Points in the plot denote the mean of each group and error bars denote 90% confidence intervals. The raw data were fitted via regression with either polynomial or linear functions. Error bands denote 90% confidence intervals for the regression estimation.

For example, within a shaking or stroking gesture, the velocity and duration were more important relative to contact area and indentation depth. With the holding gesture, nearly all attributes were of similar importance, perhaps due to this gesture's small range of motion. As for the tapping gesture, contact area was a vital attribute while one might expect vertical velocity. A potential reason is that the hand velocity measured for very fast movements, e.g., tapping/hitting in an anger expression, was not sufficient due to the camera's limited update rate of 30 frames/second.

### D. Contact attributes encode valence and arousal states

Valence and arousal ratings, between gestures, for the same message were very similar. In contrast, the ratings between messages for the same gesture were distinct (Fig. 5A). The results indicate that touchers change their contact behavior of a gesture for conveying different affective intentions. Those subtle contact changes could fine-tune the valence and arousal ratings of receivers (Fig. 5B).

First, as Fig. 5A indicates, stroking and holding gestures alike induced high valance and low arousal ratings when recognized as calm. Likewise, with the happiness message, gestures of stroking, shaking, and tapping all exhibited high valence and high arousal. Moreover, ratings for the touch-delivered messages aligned well with the ratings for the same written word stimuli [20]. These comparable emotional percepts suggest that, similar with verbal interpretation, touch may serve as a reliable channel for communicating emotion.

Second, as Fig. 5B shows, within the same gesture, changes in the contact attributes appear capable of influencing hedonic tone and intensity. In particular, with the stroking gesture the valence rating exhibits an inverted U shape relationship with contact velocity (F test, $p < 0.05$, corrected $p = 0.12$). The velocity range with the highest valence is around 10 cm/s. Moreover, the stroking gesture exhibits a higher level of valence compared with the other gestures, while its arousal rating increases as velocity increases (Spearman correlation coefficient: $\rho = 0.15$, $p = 0.078$, corrected $p = 0.16$). For the shaking gesture, a concave curve was fitted between the velocity and valence (F test, $p < 0.001$, corrected $p < 0.01$). Indentation depth displayed a significantly positive correlation with its arousal ratings ($\rho = 0.17$, $p < 0.05$, corrected $p < 0.05$). For holding and tapping gestures, greater indentation depth elicited lower valence and higher arousal (holding: valence: $\rho$ = -0.19, $p < 0.01$, corrected $p < 0.05$, arousal: $\rho = 0.23$, $p < 0.001$, corrected $p < 0.01$; tapping: valence: $\rho = -0.30$, $p < 0.0001$, corrected $p < 0.0001$, arousal: $\rho = 0.12$, $p = 0.062$, corrected $p = 0.15$).

## IV. DISCUSSION

This work shows how touchers subtly, but significantly, vary the magnitudes of their skin-to-skin contact to convey distinct social messages. Besides improving receivers' recognition of cued messages, this subtle tuning also correlates with receivers' perception of underlying valence and arousal. For instance, arousal increases with velocity for stroking, and depth for holding. More interestingly, valence is tied with velocity, here for the case of human-to-human touch, which matches the trend that has been widely reported and reproduced in the case of brushing stimuli [14], [21].

As shown in the results, the same touch gesture can deliver distinct messages if its contact attributes are subtly varied. For example, with an increase in contact velocity of around 5 cm/s, a stroking gesture is likely to be recognized as happiness instead of sympathy (Fig. 3A). With a greater indentation depth of around 1 mm, a holding gesture can be recognized as fear instead of calm (Fig. 3A). This subtle tuning in a message's recognition further quantitatively elaborates upon prior observational studies, which had indicated the same touch gesture could deliver unique emotional messages [3]–[5]. Indeed, the changes in skin's mechanics caused by those cutaneous contact attributes could elicit different responses of tactile afferents, which may underlie the discriminative and affective perception of touch [22]. For example, larger contact area may recruit more peripheral afferents [23], [24], while larger depths and forces may activate higher firing frequency [25]. Moreover, human touch interactions could be encoded in first-order neural responses of at least some afferent subtypes such as C-tactile afferents for the sensation of pleasantness [14], and single-unit Aβ afferents for distinct emotional messages [10].

Meanwhile, subtle changes in contact attributes also impact a receiver's ratings of arousal and valence. For instance, across all gestures, higher arousal was typically related to more intensive hand motions, as observed in higher velocities and greater indentation depths [3], [4], [7]. On the other hand, higher valence was associated with light contact and/or a preferred range of velocity. In agreement with prior works, this finding suggests that social touch communication goes beyond simply a participant's ability to learn a code [3], [11], [26]. Instead, it is likely that there is something more intuitive and tied to emotion in the subtlety of a touch, i.e., that these elements indeed induce and govern emotional percepts from negative to positive and from mild to intensive. Moreover, as in works with controlled brush stimuli, stroking velocity exhibits an inverted U shape relationship with the sensation of pleasantness, where the pleasant rating is low at both low (0.3 cm/s) and fast (30 cm/s) velocities but is highest around 1-10 cm/s [14], [21]. Such an inverted U shape relationship was also identified in our experiments (Fig. 5B), which to the best of our knowledge, is the first such observation in natural human touch scenarios. In addition to helping validate our measurements and findings, this point of consistency demonstrates that engineered brush stimuli are rational approximations of human-delivered strokes [12]. Furthermore, certain subtypes of peripheral afferents, i.e., SAI, SAII, Field and hair fibers, are positively correlated with stroking velocity [14]. In our setting, a similar positive trend was found between velocity and arousal (Fig. 5B), which might hint at their role in signaling of arousal percepts.

The findings herein on the importance of subtle changes may aid in the design of social haptics interfaces. Constrained by mechanical complexity, a typical design strategy in replicating human touch interactions is to mimic only a single gesture, such as hug [27], handshake [28], or stroking [29]. By focusing on slight contact changes within a gesture, future devices might render a wider range of social messages, in a more delicate fashion and via more concise form factors. Other designers have sought to represent multiple social messages using wearable actuators [9], zoomorphic robots [30], or robot manipulators [16]. Our measurements might help them in identifying and communicating more salient attributes at optimal magnitudes.